\pdfoutput=1
 \documentclass[aps, prapplied,twocolumn,showpacs,superscriptaddress,preprintnumbers, longbibliography,nofootinbib]{revtex4-2}
\usepackage[normalem]{ulem}
\usepackage{amsmath}
\usepackage{amssymb}
\usepackage{epsfig}

\usepackage{url} 

\usepackage{graphicx}
\usepackage{hyperref}
\usepackage{tabu}
\usepackage{boldline}
\usepackage{xspace}
\usepackage{slashed}
\usepackage{multirow}
\usepackage{booktabs}
\usepackage{diagbox}
\usepackage{tabularx}
\usepackage{comment}
\usepackage{floatrow}
\newfloatcommand{capbtabbox}{table}[][\FBwidth]
\usepackage{color}
\usepackage[normal]{subfigure}
\usepackage[table]{xcolor}
\usepackage{enumitem}
\usepackage[utf8]{inputenc}
\usepackage{colortbl}
\usepackage{upgreek}

\newcommand\sjt{\bgroup\markoverwith{\textcolor{blue}{\rule[0.5ex]{2pt}{0.4pt}}}\ULon}

\DeclareCaptionJustification{justified}{\justified}

\renewcommand{\arraystretch}{1.15}

\begin{document}

\preprint{}

\title{Constraints on the electron-hole pair creation energy and Fano factor below 150 eV from Compton scattering in a Skipper-CCD}

\author{A. M. Botti}
\email{abotti@fnal.gov}
\affiliation{\normalsize\it 
Fermi National Accelerator Laboratory, PO Box 500, Batavia IL, 60510, USA}

\author{S. Uemura}
\affiliation{\normalsize\it 
Fermi National Accelerator Laboratory, PO Box 500, Batavia IL, 60510, USA}

\author{G. Fernandez Moroni}
\affiliation{\normalsize\it 
Fermi National Accelerator Laboratory, PO Box 500, Batavia IL, 60510, USA}

\author{L. Barak}
\affiliation{\normalsize\it 
 School of Physics and Astronomy, 
 Tel-Aviv University, Tel-Aviv 69978, Israel}

\author{M. Cababie}
\affiliation{\normalsize\it 
Department of Physics, FCEN, University of Buenos Aires and IFIBA, CONICET, Buenos Aires, Argentina}
\affiliation{\normalsize\it 
Fermi National Accelerator Laboratory, PO Box 500, Batavia IL, 60510, USA}

 \author{R. Essig}
\affiliation{\normalsize\it 
C.N.~Yang Institute for Theoretical Physics, Stony Brook University, Stony Brook, NY 11794, USA}

\author{E. Etzion}
\affiliation{\normalsize\it 
 School of Physics and Astronomy, 
 Tel-Aviv University, Tel-Aviv 69978, Israel}

\author{D. Rodrigues}
\affiliation{\normalsize\it 
Department of Physics, FCEN, University of Buenos Aires and IFIBA, CONICET, Buenos Aires, Argentina}
\affiliation{\normalsize\it 
Fermi National Accelerator Laboratory, PO Box 500, Batavia IL, 60510, USA}

\author{N. Saffold}
\affiliation{\normalsize\it 
Fermi National Accelerator Laboratory, PO Box 500, Batavia IL, 60510, USA}

\author{M. Sofo Haro}
\affiliation{\normalsize\it  
Fermi National Accelerator Laboratory, PO Box 500, Batavia IL, 60510, USA}
\affiliation{\normalsize\it  Centro At\'omico Bariloche, CNEA/CONICET/IB, Bariloche, Argentina.}

\author{J. Tiffenberg}
\affiliation{\normalsize\it 
Fermi National Accelerator Laboratory, PO Box 500, Batavia IL, 60510, USA}

\author{T. Volansky}
\affiliation{\normalsize\it 
 School of Physics and Astronomy,   
 Tel-Aviv University, Tel-Aviv 69978, Israel}
 
\begin{abstract}

Fully-depleted thick silicon Skipper-charge-coupled devices (Skipper-CCDs) are an important technology to probe neutrino and light-dark-matter interactions due to their sub-electron read-out noise. However, the successful search for rare neutrino or dark-matter events requires the signal and all backgrounds to be fully characterized. In particular, a measurement of the electron-hole pair creation energy below 150\,eV and the Fano factor are necessary for characterizing the dark matter and neutrino signals. Moreover, photons from background radiation may Compton scatter in the silicon bulk, producing events that can mimic a dark matter or neutrino signal. We present a measurement of the Compton spectrum using a Skipper-CCD and a $^{241}$Am source.  With these data, we estimate the electron-hole pair-creation energy to be $\left(3.71 \pm 0.08\right)$\,eV at 130\,K in the energy range between 99.3~eV and 150~eV. By measuring the widths of the steps at 99.3~eV and 150~eV in the Compton spectrum, we introduce a novel technique to measure the Fano factor, setting an upper limit of 0.31 at 90\% C.L. These results prove the potential of Skipper-CCDs to characterize the Compton spectrum and to measure precisely the Fano factor and electron-hole pair creation energy below 150\,eV. 
\end{abstract}

\maketitle

\section{Compton scattering in Silicon}\label{sec:intro}

Thick fully-depleted Charge-Coupled Devices (CCD)~\cite{Smith2010, Holland:2003} built with high-resistivity silicon have become one of the most promising technologies to search for dark matter and neutrino-nucleus scatterings~\cite{Aguilar-Arevalo:2019wdi, CONNIE_2019}. By fully characterizing the background at the relevant energy range~\cite{DAMICBGD}, the DAMIC experiment used CCDs to set world-leading constraints on dark matter with masses near the GeV scale~\cite{damic_2020}, while the CONNIE experiment, based on the same technology, has set the strongest constraints on coherent neutrino-nucleus scattering at nuclear reactors in the energy range between 1 and 10\,MeV~\cite{CONNIE_2020}.

The recently developed Skipper-CCDs have allowed further progress, significantly extending dark-matter detection capabilities to sub-GeV masses. These state-of-the-art CCDs enable multiple non-destructive readouts of each pixel, thereby significantly reducing the readout noise by more than an order of magnitude,  resulting in single-electron sensitivity~\cite{Janesick1990,Wen1974, Tiffenberg:2017aac}. Several ongoing and upcoming experiments utilize Skipper-CCDs~\cite{Sensei2020,DAMIC-M,Oscura_Loi,violeta} with SENSEI already producing world-leading limits on dark matter in the eV-to-keV and MeV-to-GeV mass ranges~\cite{Crisler:2018gci,Abramoff:2019dfb,Sensei2020}.


The improved sensitivity to low-energy dark matter and neutrino interactions must be accompanied by a better understanding of low-energy backgrounds. Compton scattering constitutes an important potential background, as it can cause a high-energy photon to deposit a small fraction of its energy in the silicon bulk, thereby mimicking a dark matter or neutrino signals. Characterizing the Compton scattering spectrum at low energy is therefore of utmost importance for reducing backgrounds and identifying electron recoils due to interactions with light-dark-matter particles or neutrinos.

For free electrons, the Klein-Nishina formula~\cite{kleinnishina} describes the differential cross-section for photon scattering as a function of the energy of the incident radiation and the scattering angle. The maximal energy deposited in the interaction, known as the Compton edge, is obtained for backward scattering.

For bound electrons, the discrete energy levels must be taken into account and the cross-section is described by the Relativistic Impulse Approximation~\cite{Impulse, ImpulseErratum, Impulse2}, which results in jumps in the interaction probabilities. Theoretically, as the energy transfer increases past each step, the number of electrons available for the scattering increases, which translates to a proportional increase in the interaction rate. A summary of the silicon atomic shells is given in Table~\ref{table1}.


\begin{table}[t]

\centering
\renewcommand{\arraystretch}{1.25}
\begin{tabular}{l c c c c}
\toprule
Shell & $n$ & $\ell$ & Energy (eV) & Electrons\\
\midrule

K & 1 & 0 & 1839 & 2\\
L$_1$ & 2 & 0 & 150 & 2\\
L$_{2,3}$ & 2 & 1 & 99.3 & 6\\
Valence & 3 & - & 1.12 & 4\\
\bottomrule

\end{tabular}
\caption{Silicon atomic shells. As the energy transfer increases, an increase in the interaction rate is expected proportional to the number of additional electrons that can be ionized. The valence shell, corresponding to the smallest binding energy, is not discussed in this work.}
\label{table1}
\end{table}

The Compton spectrum above 50~eV was previously measured using traditional CCDs~\cite{comptonCCD}; however, the uncertainties in the energy estimation were dominated by the detector readout noise. Skipper-CCDs allow one to significantly reduce this noise, down to the quantum limit in which the measured spectrum is dominated by the Fano noise~\cite{Rodrigues2020} associated with the fluctuations around the average number of ionized electrons; these fluctuations smear the theoretically sharp Compton steps. With Skipper-CCDs, it is possible not only to measure the Compton spectrum with unprecedented precision but also to obtain measurements of the electron-hole pair creation energy and the Fano factor (defined as the ratio of the variance to the mean ionization). Both of these quantities need to be known in order to fully characterize a dark matter or neutrino signal. The Skipper-CCD potential to characterize the Fano noise was already demonstrated in previous work, in which the most precise measurement at about 5.9\,keV was achieved using a $^{55}$Fe source~\cite{Rodrigues2020}.

Characterizing the Compton spectrum, electron-hole pair creation energy, and Fano factor is one of the main challenges in understanding the background for measurements at energies below 100\,eV, and modeling interactions between new particles and silicon electrons or nuclei~\cite{ionizationyield}. Below, we present a first measurement of the Compton spectrum using a Skipper-CCD operated with sub-electron resolution and irradiated with a $^{241}$Am radioactive source.


\section{Measuring \texorpdfstring{$\gamma$}{gamma}-rays with a Skipper-CCD}\label{sec:setup}

\begin{figure}
\includegraphics[trim={0.0cm 0.0cm 0.0cm 0.0cm},clip,width=0.8\textwidth]{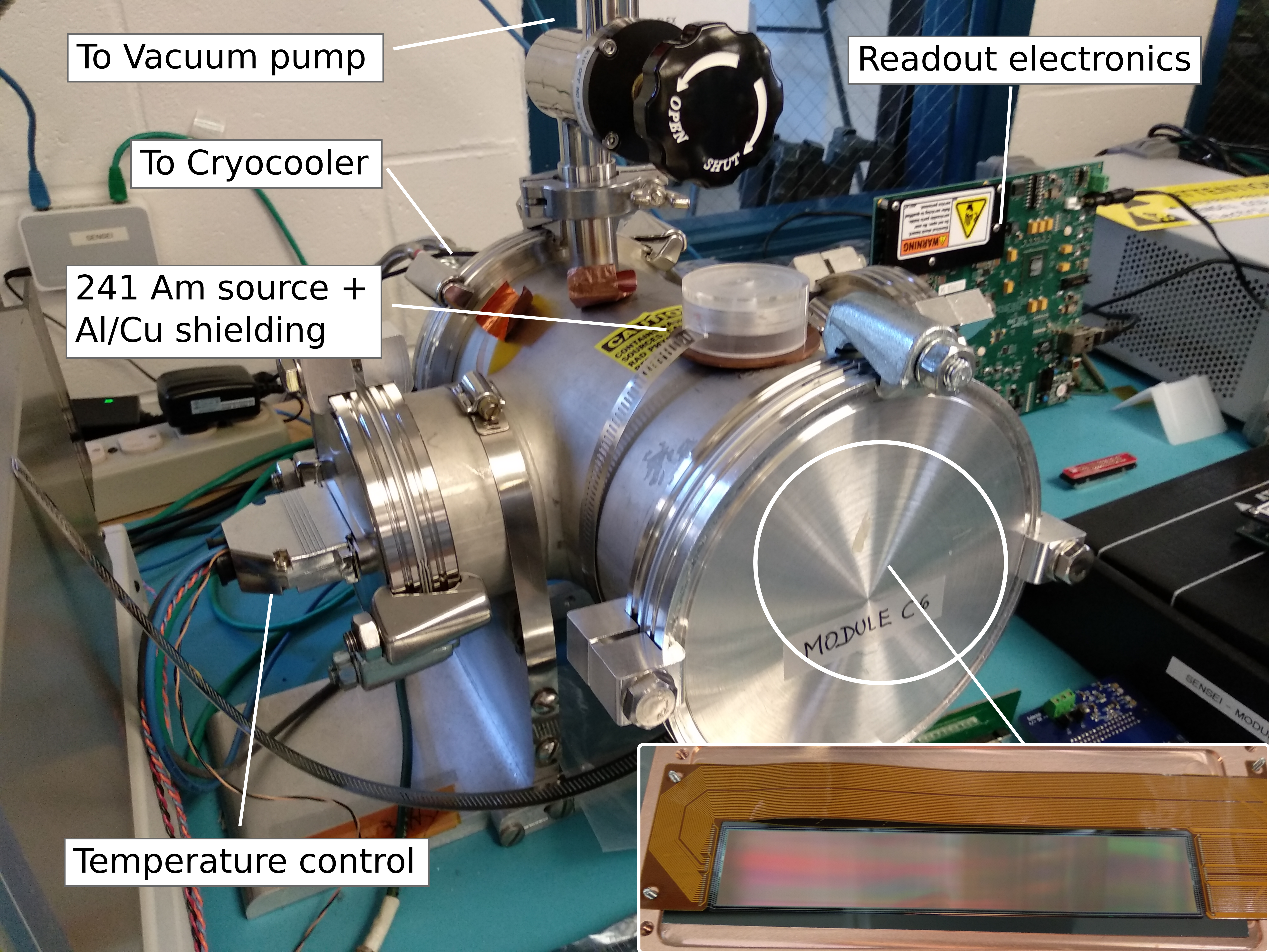}
\caption{Laboratory setup for measuring the Compton spectrum using a $^{241}$Am radioactive source. A Skipper-CCD with 1.9\,g of active mass is deployed in a copper box and installed inside an aluminum vessel for shielding. The vessel is connected to a vacuum pump, a cryocooler, and a temperature controller to operate the CCD at 130\,K. The readout electronics consists of a low-threshold acquisition board specifically designed to operate these devices.}\label{fig:setup}
\end{figure}

A picture of the experimental setup used in this work is presented in Fig.~\ref{fig:setup}. We utilized a science grade Skipper-CCD~\cite{Sensei2020} developed by the Microsystems Laboratory at LBNL and fabricated at Teledyne DALSA Semiconductor. The CCD is built with 1.9~g of high-resistivity silicon (about 20k$\Omega$-cm) divided in four quadrants, each with 3072$\times$512 pixels of volume $15 \upmu{\rm m}\times 15 \upmu{\rm m} \times 675 \upmu{\rm m}$. The CCD was placed in a copper box and deployed at Fermilab's Silicon Detector Facility (SiDet) in an aluminum vacuum vessel to shield it from environmental radiation and provide thermal isolation. On top of the vessel, a $^{241}$Am radioactive source, with an emission peak at 59.5~keV~\cite{TabRad_v5},\footnote{We note that even though the $^{241}$Am also emits $\gamma$-rays with significant probability at energies below 26.3\,keV, they are strongly attenuated by the shield. The flux of photons from the $^{241}$Am source can be considered monoenergetic for all practical purposes of this analysis} was installed with extra aluminum and copper shielding to control the rate of $\gamma$-rays reaching the CCD. Using a cryocooler and a temperature controller, the CCD was operated at 130~K and with a bias voltage of 70\,V. The readout electronics consisted of a low-threshold acquisition board configured to sample the charge in each pixel 200 times, resulting in a readout noise of about 0.22~electrons per pixel; a full description of the electronics is available in~\cite{lta}. 

To enhance the acquisition speed, we only sampled the part of the CCD closer to the radioactive source where the highest density of $\gamma$-ray hits was found; this corresponds to the first 1575 columns in each direction, as measured from the center of the CCD. Furthermore, we group the CCD pixels by bins of 10 in the direction perpendicular to the serial register (``SR''), which is the CCD readout structure. This technique allowed us to reduce the acquisition time by a factor of ten while conserving the full spatial resolution in the direction parallel to the SR, which, as will be explained in Sec.~\ref{sec:method}, eases the differentiation between $\gamma$-ray events and background. A total of 3200 images were acquired, each with a readout time of about 17~minutes. The $\gamma$-ray images contained a higher rate of single-electron events, about $8 \times 10^{-2}\,\mathrm{e^-/pix/day}$, with respect to previous measurements at the surface~\cite{backgroundAtSurface}. This is expected due to the high-energy events producing a halo of single electrons around the $\gamma$-ray hit~\cite{Sensei2020,Du:2020ldo}, and the infrared photons reaching the sensor through the copper box window (designed for measurements not related to this work). In addition, the CCD was operated in a continuous readout mode, and the clocks were optimized to maximize the charge transport efficiency needed to move thousands of electrons from pixel to pixel; this leads to a higher rate of single-electron events as reported in~\cite{SEE}.   






\section{Data selection and spectrum reconstruction}\label{sec:method}

From the whole data set (3200 images), we used the two quadrants that performed better in terms of noise and rate of single-electron events. We also removed the first 50 and last 25 columns and the first 10 and last 30 rows from each of the quadrants to avoid border effects. In addition, we removed from the analysis those images that presented a statistically unlikely high level of single-electrons; this is about 0.3\% from the whole dataset. 

\begin{figure}
\includegraphics[trim={0.0cm 0.0cm 0.0cm 0.0cm},clip,width=1.0\textwidth]{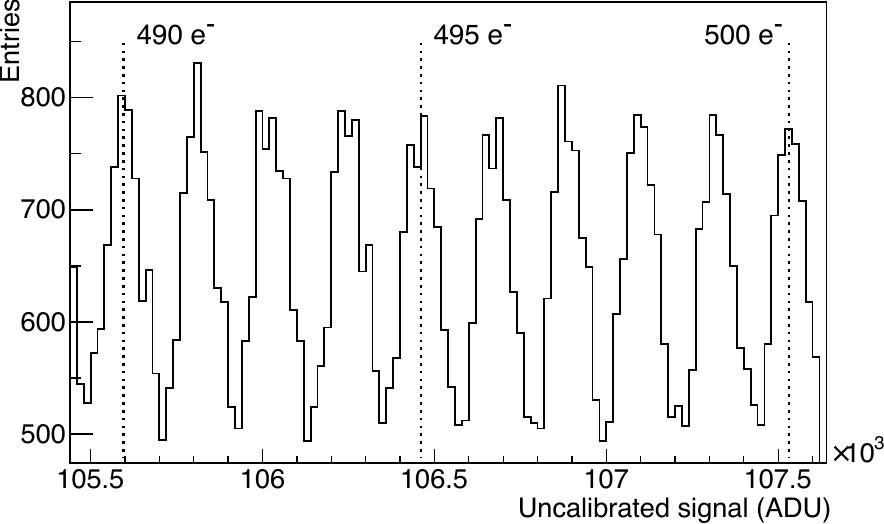}
\caption{Uncalibrated signal level by means of analog-to-digital converter units (ADUs) for collected charges between 490 and 500 electrons in one pixel. Each peak corresponds to a specific number of electrons and can be distinguished due to the sub-electron readout noise achieved with Skipper-CCDs. }\label{fig:calib}
\end{figure}

Due to the Skipper-CCD sub-electron resolution, the signal level in each pixel is easily converted to a number of electrons by reading off the peak number of the measured charge spectrum. In Fig.~\ref{fig:calib}, we show the uncalibrated spectrum in the 490 to 500 electron range; the correspondence between signal level and a number of electrons can be directly inferred. For the data used in this paper, we find an uncertainty of $0.22$ electrons in this conversion.  A full description of the Skipper-CCD calibration for higher-energy interactions is presented in~\cite{Rodrigues2020}.

After calibration, a clustering algorithm is used for each image to reconstruct events produced by particle interactions. The algorithm searches for pixels with a charge above 0.6\,electrons and groups together all neighboring pixels that match this condition as one event. We then applied quality cuts based on the pixel cluster geometry to separate the $\gamma$-ray clusters from the background.
Since blocks of 10 pixels are binned into a ``super-pixel'' during readout, the probability of having accidental coincidences between uncorrelated background electrons increases. This results in extended clusters of several 1- or 2-electron super-pixels, with a charge density significantly lower than that of $\gamma$-rays. The first quality cut is aimed at removing this background by rejecting all clusters in which the average charge per super-pixel is lower than five electrons. The second quality cut also constrains the spatial distribution of the cluster charge. Upon an interaction in the silicon bulk, charges drift toward the surface of the CCD, spreading across several pixels. The resulting cluster geometry around the point of interaction, which is assumed to be at the pixel with maximal charge, is dictated by a normal distribution. The variance depends on how far from the surface the interaction took place except for events with a small overall charge, where the quantization noise dominates the variance. Conversely, when an interaction takes place in the inactive silicon around the SR~\cite{backgroundAtSurface}, the charges produced in the undepleted silicon may diffuse into the collection region of the SR pixels. These events tend to present a distinct geometry from those produced in the bulk, as these are spread with a larger variance in only one direction. Below twenty electrons, the SR and bulk events are statistically indistinguishable partly since the geometry in the binning direction does not contribute to the discrimination. We set an uppercut in the standard deviation of the binned direction, perpendicular to the SR, of 0.5 pixels to remove merged clusters. Furthermore, we only select events with a standard deviation in the direction parallel to the SR ($\sigma_\parallel$), where the full spatial information was preserved, between 0.3 and 1.0 pixel.

\begin{figure}
\includegraphics[trim={0.0cm 0.0cm 0.0cm 0.0cm},clip,width=1.0\textwidth]{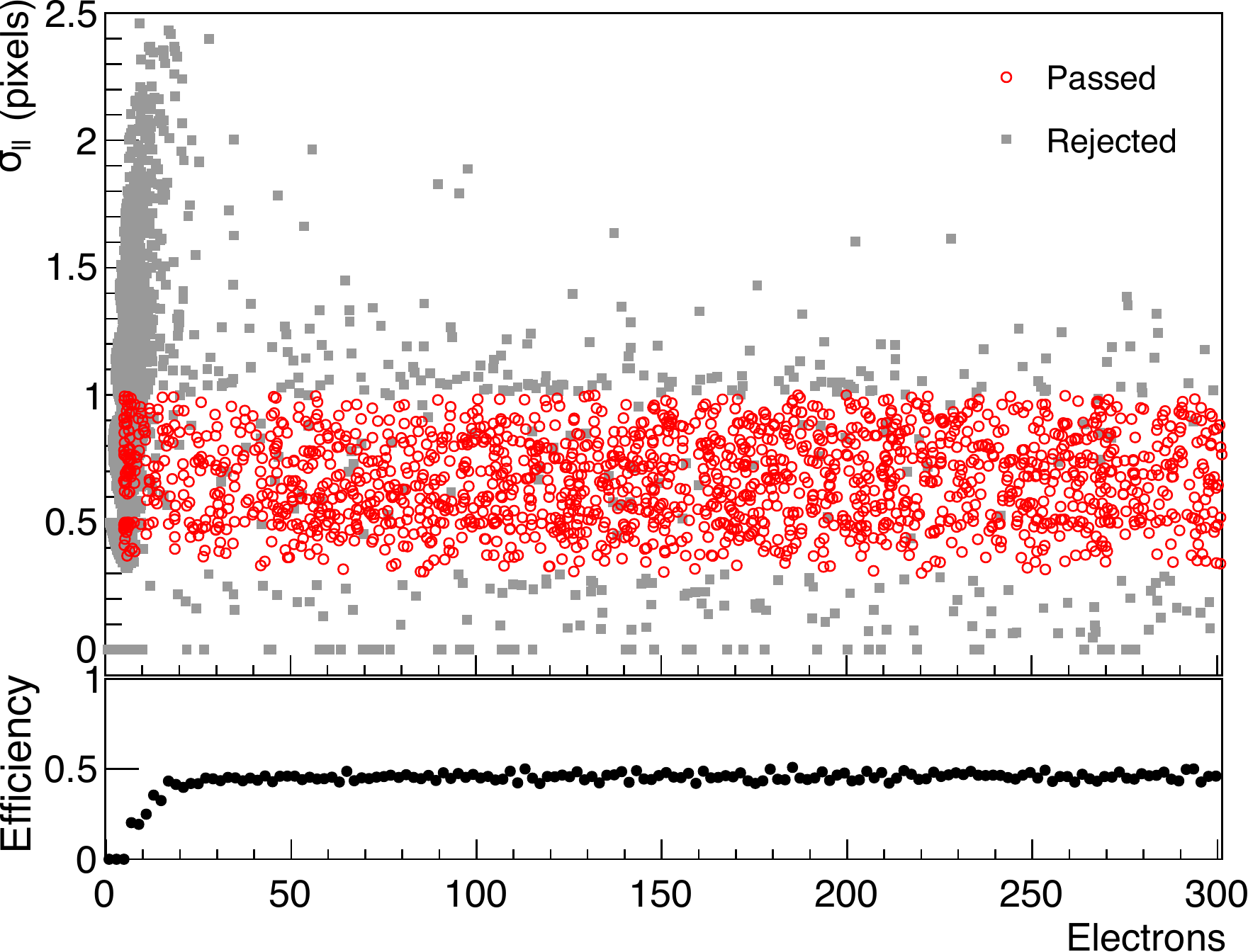}
\caption{(Top) standard deviation of the cluster charge in the unbinned direction as a function of the cluster charge. Events rejected by the quality cuts (gray squares) correspond mostly to serial register hits, and events that pass the quality cuts (red circles) correspond mostly to interactions in the silicon bulk. (Bottom) Monte Carlo estimate of the event reconstruction and selection efficiency as a function of the number of electrons.}\label{fig:varianceEff} 
\end{figure}

In the top panel of Fig.~\ref{fig:varianceEff}, we present $\sigma_\parallel$ as a function of the number of electrons for events that were rejected by the quality cuts (gray squares) and those that passed (red circles). Events produced in the bulk have a reconstructed variance that does not depend on the number of electrons except when this number is rather small (about 20 electrons).
In contrast, SR hits can be identified as those with charge below 50 electrons and high $\sigma_\parallel$. Charge at the edge of a low-charge SR hit tends to be disconnected from the cluster and lost in reconstruction, leading to a correlation between charge and width. SR events with less than 20 electrons are therefore indistinguishable from those produced in the silicon bulk.

The above selection criteria significantly reject background events without introducing any biases down to 20~electrons, as we illustrate in the bottom panel of Fig.~\ref{fig:varianceEff}. To verify that our quality cuts are not distorting the Compton spectrum, we use a Monte Carlo simulation based on a diffusion model~\cite{SofoDiffusion, Sensei2020Sup} that describes how the electrons produced in the silicon bulk propagate towards the CCD surface. If a $\gamma$-ray interacts close to the CCD surface, electrons will not spread much, resulting in small clusters, while an interaction occurring in the back will typically result in bigger clusters. In this sense, the variance cuts described above indirectly select events that are produced in a certain region of the bulk. However, at low energy, the maximal spread may strongly depend on the number of electrons collected regardless of where the interaction occurs. Therefore, the selection efficiency will depend on the interaction energy and may distort the measured spectrum. We verified that this is not the case for events with more than 20~electrons using simulations: for a given number of electrons, we injected simulated diffused clusters in real images and computed the probability of successfully reconstructing them after applying the quality cuts. This allowed us to obtain the selection efficiency as a function of the number of electrons for the different images. A low occupancy in the image translates to a smaller probability of having clusters piling up or merging with each other, resulting in higher efficiency, as is the case for the images without the $\gamma$-ray events. Conversely, a higher occupancy, as we have in the $\gamma$-ray images, results in lower efficiency.


\section{Reconstructed spectrum and model comparison}\label{sec:discussion}

\begin{figure*}[t]
\includegraphics{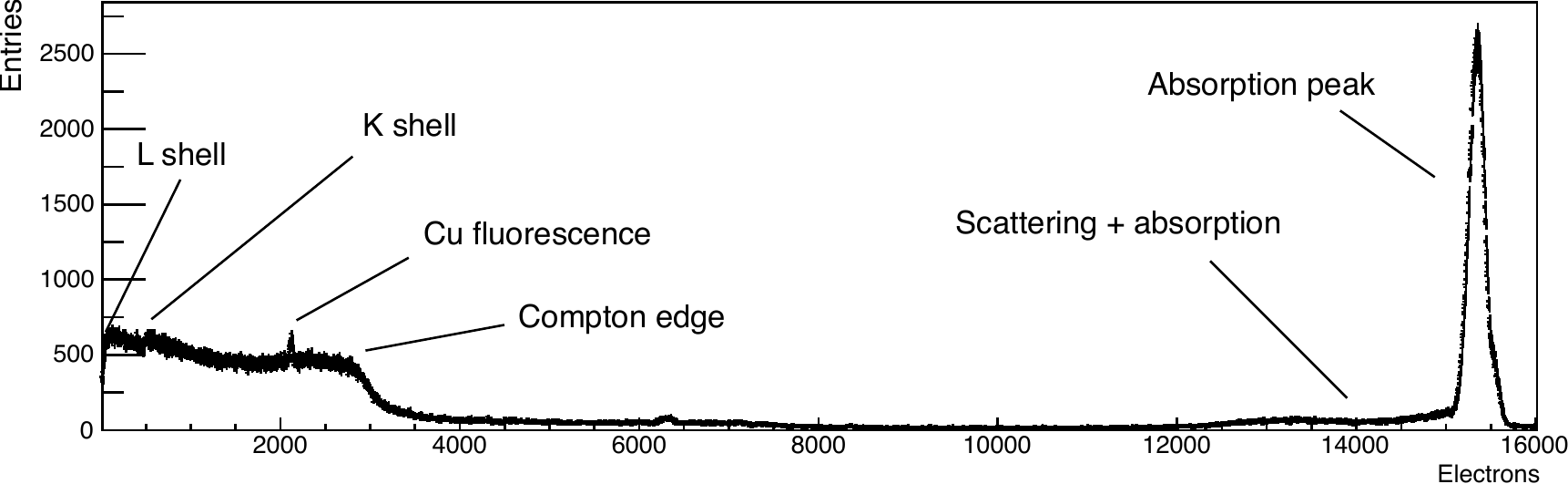}
\caption{\label{fig:spectrumNotCalin} Compton spectrum of 59.54\,keV $\gamma$-rays in silicon measured with a Skipper-CCD. The photo-absorption peak at about 15,000 electrons is observed, along with steps matching the atomic-shell energies of silicon. The Compton edge is observed at about 3000 electrons along with the photo-absorption peak shoulder at about 13,000 electrons corresponding to photons that first lost energy through Compton-scattering and were later absorbed. On the right of the absorption peak a knee is obtained, which is produced by the difference in gain between quadrants. This is expected since we only calibrated the signals up to 500 electrons. One of the copper fluorescence peaks is observed at 2,100 electrons, which is produced by $\gamma$-rays exciting the copper tray in which the CCD is deployed. Finally, between 3,500 and 8,000 electrons a small number of events that result from the pile-up of two or more $\gamma$-ray events was identified.}
\end{figure*}


For illustration, the full charge spectrum of the $\gamma$-ray images after calibration and data selection is presented in Fig.~\ref{fig:spectrumNotCalin}. The steps that stem from the atomic structure are visible at the lower end of the spectrum. The first step at 1839\,eV (about 490 electrons) corresponds to the K-shell, where two target electrons are lost; this gives a drop in the rate of Compton-scattered events of $12 / 14 = 0.86$. The first L-shell at 150\,eV also corresponds to a loss of two target electrons, giving a drop of $10 / 12 = 0.83$. After the second and third L-shell at 99.3\,eV, six additional target electrons are lost, which gives an expected drop of $4 / 10 = 0.4$. 

\begin{figure}[t]
\includegraphics[trim={0.0cm 0.0cm 0.0cm 0.0cm},clip,width=1.0\textwidth]{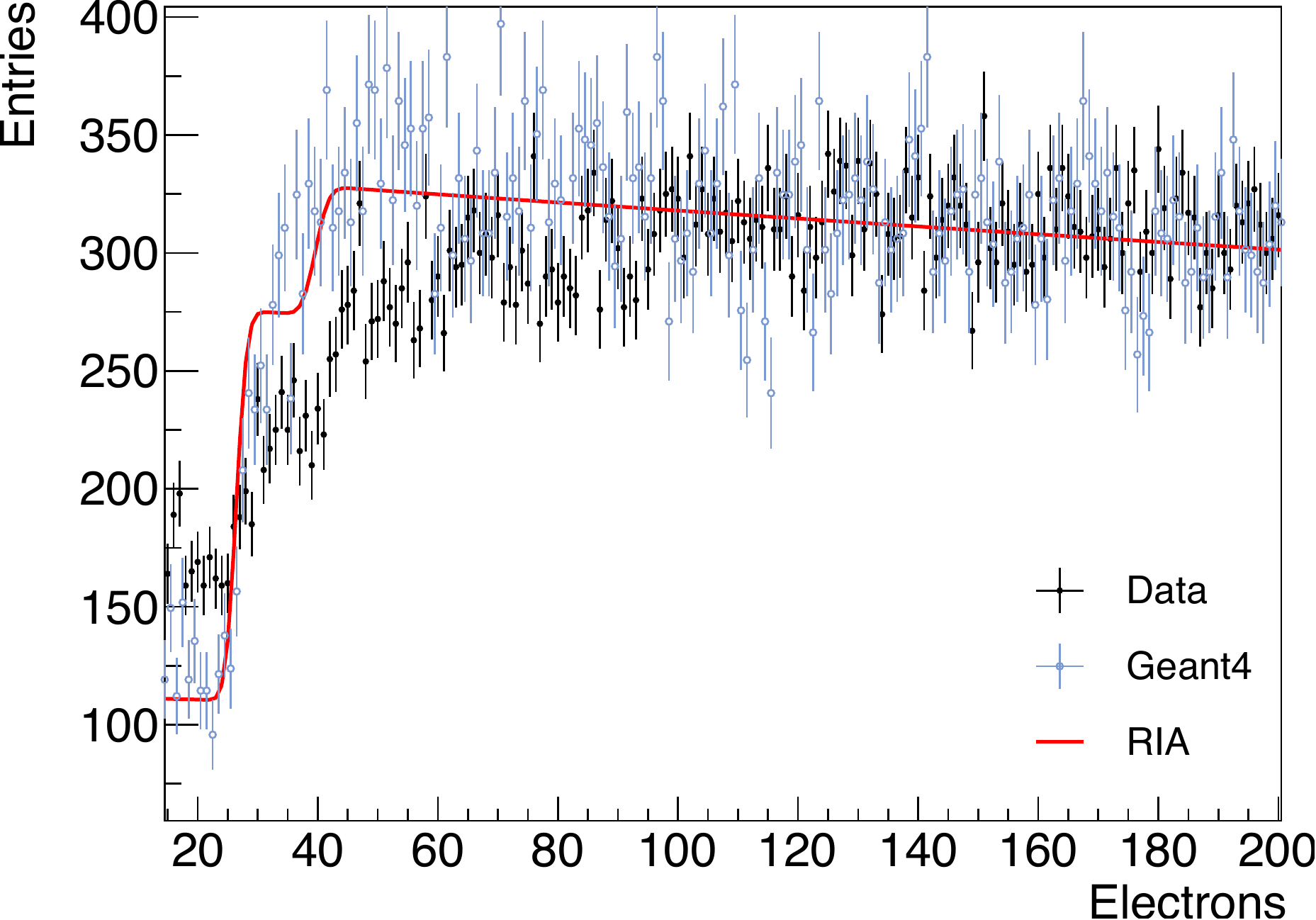}
\caption{Measured spectrum (black markers) compared to the RIA theoretical prediction (solid red line) and a GEANT4 simulated spectrum (blue markers) using an implementation of the RIA model. The RIA spectra are normalized to the data in the 100-200 electron range, far from the shell effects. Please refer to the main text for additional details.}\label{fig:theo_vs_data}
\end{figure}

Currently, the most precise and community accepted theoretical model for low-energy Compton scattering is the Relativistic Impulse Approximation (RIA) proposed by~\cite{Impulse}. To obtain the RIA-predicted spectrum we used the X-ray and atomic structure information provided by~\cite{SCHOONJANS2011776}. We included the effect of the energy shift and Fano fluctuations by convolving the theoretical spectrum with a Gaussian function with the reference parameters in Table~\ref{results}. Figure~\ref{fig:theo_vs_data} shows a significant discrepancy between the predicted curve (red line) and the measured data (black full circle), as already hinted in~\cite{comptonCCD}.

To prove that this discrepancy is not produced by edge effects arising from partial energy deposition of interactions close to the front and back of the sensor, we also performed a full GEANT4 simulation. This contains detailed information on the detector geometry (including the dead layers as described in~\cite{PhysRevApplied.15.064026}) and uses the RIA-based Penelope physics list as recommended for low-energy electromagnetic interactions\footnote{This RIA model uses a slightly different information for the atomic shells compared to~\cite{SCHOONJANS2011776} which we consider more accurate, but we do not expect a significant impact from this small discrepancy.}. We then injected the GEANT4 events in empty images, including the Fano fluctuations and modeling the charge diffusion in transport to the front of the CCD~\cite{SofoDiffusion}, processed these images through our standard processing chain, and applied the selection criteria described in Sec.\ref{sec:method}.

In Fig.~\ref{fig:theo_vs_data} we show that the numerically-computed RIA model and the GEANT4 simulation (blue empty circles) are in good agreement. We can then confirm that the detector-edge effects are not the source of the discrepancy observed in the 30-80 electron region between the measured and RIA spectra\footnote{It may be of interest for the reader that since the submission of this manuscript a new paper that discusses this discrepancy has appeared in the arXiv~\cite{https://doi.org/10.48550/arxiv.2207.00809}.}. 

\section{Extracting the electron-hole pair creation energy and Fano factor}\label{sec:fit}
As mentioned in the previous section, the spectrum steps can be described in terms of the drop in the number of target electrons at each shell energy. Since the RIA does not provide an accurate description of the data, we use a simplified model in which each step is described by a Heaviside step function positioned at the specific shell energy. Detector and statistical effects such as the readout and Fano noise introduce a Gaussian uncertainty in the energy measurements\footnote{While the Fano noise is not strictly Gaussian, the steps are at sufficiently high energy to justify a Gaussian approximation~\cite{ionizationyield}.}.

As a result, each step can be modeled as the convolution of a Heaviside step function with a Gaussian distribution that describes the effective energy resolution and includes intrinsic charge-generation fluctuations (Fano factor) and detector effects (readout noise and dark current): 
\begin{equation}
\centering
\begin{split}
 A \Theta\left(\bar{x}\right) * G\left(\bar{x}\right) + K &= A \int_{\tau}^{\infty} G(\bar{x}-\tau) \,d\tau + K\\
 &= \frac{A}{2}\left(\mathrm{Erfc}\left(\frac{\mu-\bar{x}}{\sqrt{2}\sigma}\right)\right) + K\,.
 \label{model}
\end{split}
\end{equation}
Here, $\Theta\left(\bar{x}\right)$ and $G\left(\bar{x}\right)$ are a Heaviside step and Gaussian functions
respectively, $A$ is the increase in the number of events after the step, $K$ is the number of events at the lower part of the step, $\mu$ is the position of the step, and $\bar{x} = x-b$ is the measured charge $(x)$ minus a reconstruction bias ($b$), which is produced by the merging of single electrons into the clusters. We determined this bias using simulated events on real images and obtained $b=1.8$ electrons at the first step and $b=1.9$ electrons at the second. $\sigma = \sqrt{\sigma_{RO}^2 + \sigma_{sys}^2 + Fx}$ is the energy resolution that includes contributions from the readout noise ($\sigma_{RO}$), Fano fluctuations ($Fx$), and a systematic uncertainty ($\sigma_{sys}$) produced by fluctuations in the energy estimate introduced by the high rate of single-electron events. While alternative models may provide a better description of the physics, to first order any correction can be treated as additional contributions to the energy resolution and the assumption of a Heaviside step leads to a conservative constraint on the energy resolution.

\begin{figure}[t]
\includegraphics[trim={0.0cm 0.0cm 0.0cm 0.0cm},clip,width=1.0\textwidth]{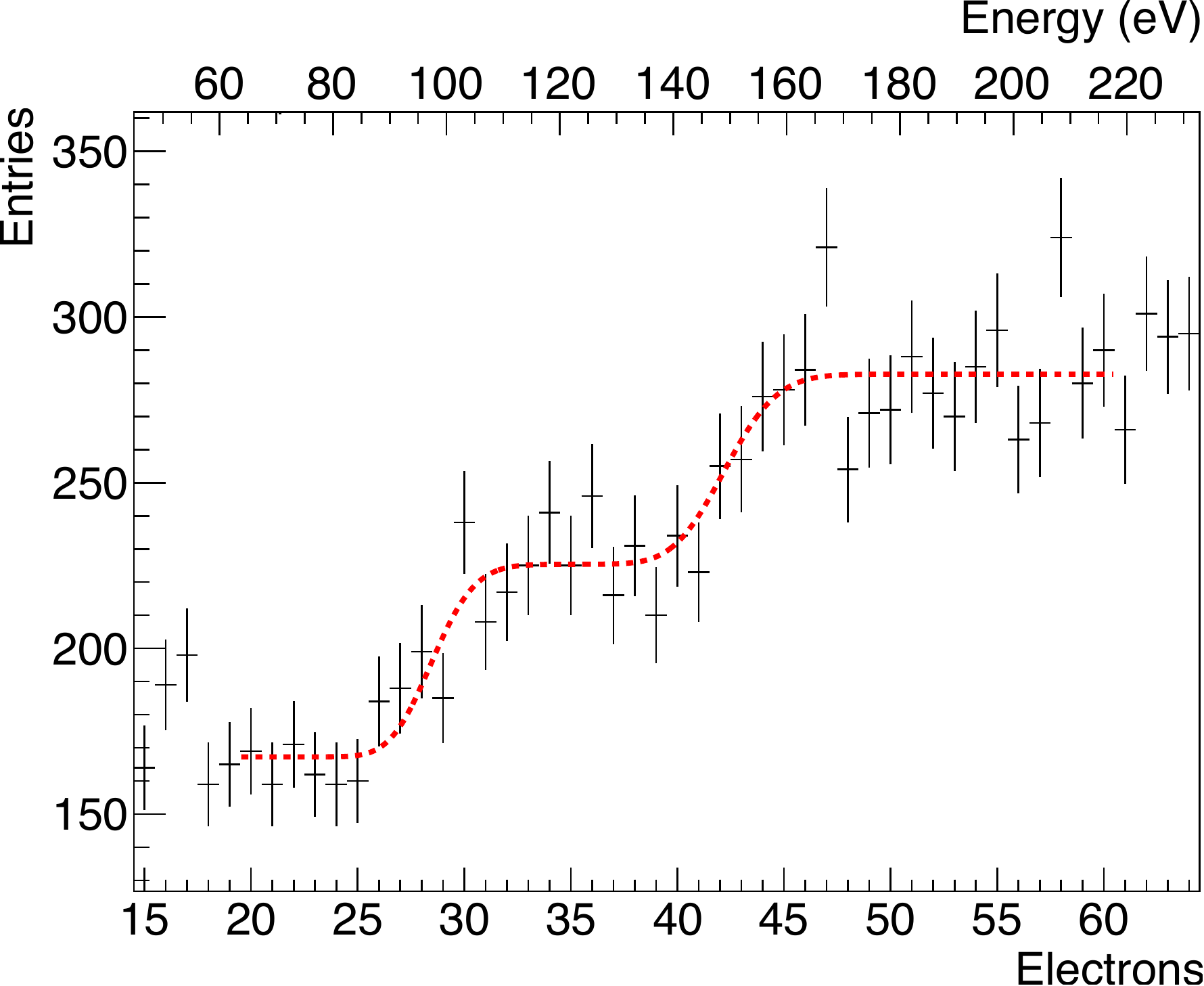}
\caption{Measured Compton steps at the L-shell energies. The red line corresponds to the fit of a phenomenological model, which consists of the convolution of Heaviside functions with Gaussian distributions whose widths correspond to the energy resolution as illustrated in~\ref{model}}\label{fig:lowESteps}
\end{figure}

In Fig.~\ref{fig:lowESteps}, we show the measured Compton spectrum (black data points) from 20 to 60~electrons (bottom axis). The steps corresponding to the L-shells (99.3~eV and 150\,eV) are observed at about 29 and 42 electrons. On the top axis, we show only for illustration the energy corresponding to each number of electrons with an average bias correction of 1.85 electrons and assuming an electron-hole pair creation energy of 3.71 eV, which, as we explain later in this section, is extracted from the data. Below 18 electrons the data is background-dominated, mainly due to the contribution of the SR events, as explained in Sec.~\ref{fig:calib}. The dashed red line shows the fitted convolution of the Heaviside step function with a Gaussian distribution. For the fit, we assume that the electron-hole pair conversion energy and Fano factor remain constant throughout the whole fitted range. Under the conservative assumption that the Fano noise is the sole contribution to the width of the steps, we obtain an upper limit on the Fano factor. We test the robustness of the model and the statistical errors on the fit results by repeating the fit on Poisson-distributed toy data samples drawn from the fitted Heaviside-Gaussian model. We also use toy data to verify that there is no systematic uncertainty associated with the fit method and the selection of the fitting range. In addition, we run a likelihood-ratio test using a single-step function as a particular case of the two-step hypothesis. We obtained a p-value of $3 \times 10^{-4}$, thus rejecting the single-step hypothesis. This result denotes an improvement from the previous measurement with traditional CCDs~\cite{comptonCCD}, where the L$_1$- and L$_{2,3}$-shell steps could not be resolved and a one-step hypothesis had to be assumed.

By fitting the position of the steps, $\mu$, we obtain an electron-hole pair creation energy of $\varepsilon_{eh} = (3.71 \pm 0.08)$\,eV. This result is similar to the one previously obtained using Skipper-CCDs at 5.9\,keV and 123\,K~\cite{Rodrigues2020}, where a pair creation energy of (3.752 $\pm$ 0.002)\,eV was reported. Even though these values depend on the interaction energy and lattice temperature and cannot be compared directly, they seem within the expectation of extrapolating to lower energy and are also consistent with other measurements performed using silicon-based technologies and semi-empirical models~\cite{ionizationyield,LECHNER1995464,FRASER1994368}. 

A precise estimate of the Fano factor, $F$, using the Heaviside-Gaussian model is hampered by the limited available statistics. Furthermore, the toy Monte Carlo model suggests that the Fano factor is the only fit parameter that does not follow a normal distribution, which makes it challenging to estimate a confidence interval. Nevertheless, if we assume that the energy resolution is solely due to Fano fluctuations, we can set an upper limit on the Fano factor of 0.31 with a 90\%~C.L.: 90\% of toy datasets generated with this value have fitted values lower than the value fitted from the data. This result is also consistent with previous measurements using Skipper-CCDs, where the Fano factor obtained was 0.119\,$\pm$\,0.002 at 5.9\,keV and 123\,K~\cite{Rodrigues2020}.

We define the step size to be the ratio between the lower and the upper part of the step; this is $\frac{K}{K+A}$, where $K$ and $A$ are extracted from Eq.~(\ref{model}). For the first step at 150\,eV (about 40 electrons) we obtain a drop of $0.80 \pm 0.02$, compatible with the expectation of 0.83. For the second step at 99.3\,eV (about 25 electrons), we find a decrease in the number of events of $0.74 \pm 0.03$, which is statistically inconsistent with the expected value of 0.4.

In this data set, we found a high rate of single-electron events, $8 \times 10^{-2}$\,e$^-$/pix/day. These electrons are not uniformly distributed along the CCD, and we do not fully characterize their contribution to the energy resolution. 
The reconstruction bias introduced by single-electron events that are coincident with the $\gamma$-ray clusters is determined using a simulation and included in the fit model. 

A full study of the discrepancy in the step size at 99.3\,eV is still in progress and will be the focus of a follow-up publication. However, a simulation shows that injecting a density of single-electron events, similar to that in the $\gamma$-ray images, to images with only serial register hits results in a wider charge spectrum due to the merging of clusters. This explains the measured shallower step: background events in the 20 to 30 electron region yield the discrepancy between data and expectations. Furthermore, preliminary results from a new measurement show that the step size is recovered after reducing the single-electron rate and X-ray occupancy in the images.

To obtain a precision measurement it is of the utmost importance to reduce the density of single-electron events in the images, which in addition to the effects above also contributes to fluctuations in the reconstructed energy of the clusters. This understanding is one of the main reasons to improve the data quality in future work. In particular, we will aim at reducing the image occupancy to control the number of electrons in the halo of high-energy events~\cite{Sensei2020Sup} and will further optimize the clock voltages to reduce the spurious charge~\cite{SEE}. Furthermore, we intend to increase the size of the data set, and thus reduce the statistical uncertainty.



\section{Summary and outlook}\label{sec:outllok}

In this work, we presented a measurement of the Compton spectrum for 59.54\,keV $\gamma$-rays interacting in silicon using a science-grade Skipper-CCD operated at 130\,K. The low-energy spectral steps corresponding to the atomic energy shells were observed, and in particular, the two steps matching the L-shells at energies of 99.3\,eV and 150\,eV were distinguished. We compared our data with two implementations of the RIA model proving that this is not an accurate description of the Compton spectrum below 100 electrons. We provided a phenomenological model that describes the resulting spectrum using a convolution of a Heaviside step function with a Gaussian distribution, which allowed us to study the impact of the detector's energy resolution on the shape of the Compton steps. Results after fitting this model to the measured spectrum are presented in Table~\ref{results}, in tandem with reference values from theoretical expectations and previous work with Skipper-CCDs at 5.9\,keV~\cite{Rodrigues2020}.

\begin{table}[h]

\centering
\renewcommand{\arraystretch}{1.25}
\begin{tabular}{c c c}
\toprule
Parameter & Result & Reference\\
\midrule
$\varepsilon_{eh}$ (eV) & 3.71 $\pm$ 0.08 & 3.75~\cite{Rodrigues2020}\\
$F$ & $<$ 0.31 (90\% c.l.) & 0.12~\cite{Rodrigues2020}\\
150\,eV Step & 0.80 $\pm$ 0.02 & 0.83 \\
99.3\,eV Step & 0.74 $\pm$ 0.03 & 0.40 \\
\bottomrule

\end{tabular}
\caption{Electron-hole pair creation energy ($\varepsilon_{eh}$), Fano factor ($F$), and size of the steps obtained after fitting the Compton spectrum between 70 and 200\,eV with the convolution of a Heaviside step function with a Gaussian distribution. Previous results with Skipper-CCD~\cite{Rodrigues2020} and theoretical expectations are shown for comparison.}
\label{results}
\end{table}

We used this measurement to set novel constraints on the electron-hole pair creation energy ($\varepsilon_{eh}$) and Fano factor ($F$) at the energies of the silicon atomic L-shells: 99.3\,eV and 150\,eV. Our results are consistent with previous work using the same technology but measured at higher energy and different operating temperature. Theoretical expectations and other measurements with silicon-based technologies are also consistent with our results~\cite{ionizationyield,LECHNER1995464,FRASER1994368}. Finally, the size of the step corresponding to the second and third L-shells (99.3\,eV) is not consistent with the simple theoretical expectation that only considers the change in the number of available electronic targets on which photons may scatter. A detailed study to understand the nature of this discrepancy is planned for future work. A plausible hypothesis of this discrepancy is the high rate of single-electron events that contributes to the measured cluster energy. Controlling the rate of single electrons by reducing the image occupancy and optimizing the CCD operation~\cite{SEE, smartSkipper} are the next steps towards a precision measurement of the electron-hole pair creation energy and Fano factor below 150\,eV using Compton scattering in Skipper-CCDs.



\begin{acknowledgments}

This work was supported by Fermilab under DOE Contract No.\ DE-AC02-07CH11359. 
The CCD development work was supported in part by the Director, Office of Science, of the DOE under No.~DE-AC02-05CH11231. 
We are grateful for the support of the Heising-Simons Foundation under Grant No.~79921. 
R.E.~acknowledges support from the Simons Investigator in Physics Award 623940 and the US-Israel Binational Science Foundation Grant No.~2020220.
TV is supported, in part, by
the Israel Science Foundation (grant No.~1862/21), by
the Binational Science Foundation (grant No.~2020220)
and by the European Research Council (ERC) under the
EU Horizon 2020 Programme (ERC-CoG-2015 - Proposal
n.~682676 LDMThExp). This manuscript has been authored by Fermi Research Alliance, LLC under Contract No. DE-AC02-07CH11359 with the U.S.~Department of Energy, Office of Science, Office of High Energy Physics. The United States Government retains and the publisher, by accepting the article for publication, acknowledges that the United States Government retains a non-exclusive, paid-up, irrevocable, world-wide license to publish or reproduce the published form of this manuscript, or allow others to do so, for United States Government purposes.
\end{acknowledgments}



\bibliography{Compton.bib}

\end{document}